\documentclass{appolb}
\usepackage{graphicx}
\usepackage{authblk}
\usepackage{cite}

\begin{document}
\sloppy
\title{Potential of the J-PET detector for studies of discrete symmetries 
       in decays of positronium atom - a purely leptonic system%
       %
}
\author{
P.~Moskal$^{a}$,
D.~Alfs$^{a}$, 
T.~Bednarski$^{a}$, 
P.~Bia\l as$^{a}$, 
E.~Czerwi\'nski$^{a}$, 
C.~Curceanu$^{b}$,
A.~Gajos$^{a}$, 
B.~G\l owacz$^{a}$, 
M.~Gorgol$^{c}$,
B.~C.~Hiesmayr$^{d}$,
B.~Jasi\'nska$^{c}$, 
D.~Kami\'nska$^{a}$, 
G.~Korcyl$^{a}$, 
P.~Kowalski$^{e}$, 
T.~Kozik$^{a}$, 
W.~Krzemie\'n$^{f}$, 
N.~Krawczyk$^{a}$, 
E.~Kubicz$^{a}$, 
M.~Mohammed$^{a}$, 
Sz.~Nied\'zwiecki$^{a}$,  
M.~Pawlik-Nied\'zwiecka$^{a}$, 
L.~Raczy\'nski$^{e}$, 
Z.~Rudy$^{a}$, 
M.~Silarski$^{a}$, 
A.~Wieczorek$^{a,g}$, 
W.~Wi\'slicki$^{e}$, 
M.~Zieli\'nski$^{a}$ 
}

\affil{
       $^{a}$Faculty of Physics, Astronomy and Applied Computer Science, Jagiellonian University, 30-348 Cracow, Poland\\
       $^{b}$INFN, Laboratori Nazionali di Frascati, CP 13, Via E. Fermi 40, I-00044, Frascati, Italy\\
       $^{c}$Department of Nuclear Methods, Institute of Physics, Maria Curie-Sk\l odowska University, 20-031 Lublin, Poland\\
       $^{d}$ Faculty of Physics, University of Vienna, Boltzmanngasse 5, 1090 Vienna, Austria\\
       $^{e}$\'Swierk Computing Center, National Center for Nuclear Research, 05-400 Otwock-\'Swierk, Poland\\
       $^{f}$High Energy Physics Division, National Center for Nuclear Research, 05-400 Otwock-\'Swierk, Poland\\
       $^{g}$Institute of Metallurgy and Materials Science of Polish Academy of Sciences, 30-059 Cracow, Poland\\
     }

\maketitle
\begin{abstract}
The Jagiellonian Positron Emission Tomograph (J-PET) 
was constructed as a prototype of the cost-effective scanner for the 
simultaneous metabolic imaging of the whole human body.
Being optimized for the detection 
of photons from the electron-positron annihilation
with high time- and high angular-resolution,
it constitutes a multi-purpose detector providing new opportunities 
for studying the decays of positronium atoms.

Positronium is the lightest purely leptonic object decaying into photons. 
As an atom bound by a central potential 
it is a parity eigenstate, and as an atom built out of an electron and an anti-electron
it is an eigenstate of the charge conjugation operator. 
Therefore, the positronium is a unique laboratory to study discrete symmetries 
whose precision is limited in principle by the effects due 
to the weak interactions expected at the level of ($\sim\!10^{-14}$) 
and photon-photon interactions expected at the level of~($\sim\!10^{-9}$). 

The J-PET detector 
enables to perform tests of discrete symmetries in the leptonic sector
via the determination of the expectation values of the discrete-symmetries-odd operators, 
which may be constructed 
from the spin of ortho-positronium atom and
the momenta and polarization vectors of photons originating from its annihilation. 
In this article we present the potential of the J-PET detector to test the 
C, CP, T and CPT symmetries in the decays of positronium atoms. 
\end{abstract}
  
\section{Introduction}
If Nature was utterly symmetric the matter would not exist. 
Thus, we owe our existence to the asymmetry between matter and anti-matter 
which must have appeared at the early stage of the evolution of the Universe. 
Surprisingly though, processes driven by the gravitational, 
electromagnetic and strong interactions up to now appear to be symmetric 
with respect to reflection in space (P), 
reversal in time (T) 
and charge conjugation (C).
So far violations of these symmetries were observed only 
in processes governed by the weak interaction. 
Breaking of C and CP symmetries is a necessary condition for baryogenesis to occur 
(for generation of asymmetry between baryons and anti-baryons), 
independently of the particular mechanism~\cite{m1}.  
Yet, the excess of matter over anti-matter which we observe in the current Universe 
is by about nine orders of magnitude too large with respect to the theoretical 
estimations based on the presently known sources of the discrete symmetries violations~\cite{m2}. 
This huge discrepancy remains one of the greatest puzzles in physics and cosmology. 

The origin of the excess of matter over anti-matter in the Universe may also be explained 
by the 
leptogenesis~\cite{m3} 
which is referred to as the hypothesis explaining the
existence of the matter in the Universe by the appearance of lepton-antilepton asymmetry 
in the early stage of its evolution. The present theory of leptogenesis postulates that 
lepton-antilepton asymmetry was generated in decays of the hypothetically heavy right-handed 
neutrinos and it was converted to the baryon-antibaryon asymmetry via interaction 
of hypothetical sphalerons~\cite{m4}. 
However, the existence of both mentioned particles was not confirmed up to now.
Interestingly, though the matter which we know is made of quarks and leptons, 
violation of CP and T symmetries have been observed only for systems including quarks, 
and it has not yet been discovered in any processes involving purely leptonic matter. 
Such studies are challenging and the best so far performed experiments 
with positronium atoms excluded violation of discrete symmetries as CP, T or CPT only 
at the level of about 0.3\%~\cite{m35,m47},
which is many orders of magnitude less precise than the accuracies achieved in the quark sector. 

By using the J-PET detector we plan to investigate symmetries in the decays of positronium ($e^+e^-$) 
which is composed of lepton and anti-lepton constituting the simplest atom made of matter and anti-matter. 

This article is organized as follows:
first we describe the basic properties of the positronium atom,
next we present the state of the art and detailed justification for C, CP, T and CPT symmetries breaking. 
Further on, 
we explain the principle of operation
of the J-PET tomograph including introduction 
of methods developed for the creation of spin- and tensor polarized positronium atoms.
Next, we describe the techniques for  monitoring of the linear 
and tensor polarization of positronium atoms,
as well as the method for the determination of photons polarization. 
This description is followed by an estimate of the limitations for the discussed studies
due to the instrumental and physical backgrounds.
Finally, we introduce odd-symmetric operators (constructed from ortho-positronium spin,
as well as momentum and polarization vectors of photons)   
which will be used for the tests of the discrete symmetries. 

It is important to stress that the tests of symmetries with positronium described hereafter in this article
are independent of and thus not limited by the results of the tests for T and CP invariance 
in muon decays (e.g. $\mu \to e \nu \bar{\nu}$) 
and by the searches of the electric dipole moments~\cite{m13}.

\section{Positronium}
Positronium~\cite{m9} is simultaneously an atom and an anti-atom. 
It is build out of an electron and an anti-electron, and thus it is a purely leptonic object. 
Similarly as atoms bound by a central potential it is an eigenstate of the parity operator P. 
In addition, unlike the ordinary atoms, but similarly to the flavor neutral mesons, 
the positronium is symmetric under the exchange of particles into antiparticles and so it is 
also an eigenstate of the charge conjugation operator C and thus an eigenstate of the CP operator. 
It is the simplest atomic system with charge conjugation eigenstates~\cite{m10}. 
This makes it ideal for studies of the discrete symmetries in Nature. 

Positronium, similarly to the hydrogen atom, 
in the ground sub-states with orbital angular momentum L~=~0, 
is formed in a singlet state of the anti-parallel spins orientation, 
the so-called para-positronium (p-Ps), 
or in a triplet state of parallel spin orientation, 
the so-called ortho-positronium (o-Ps). 
Due to the symmetry of charge conjugation, 
p-Ps undergoes annihilation with emission of an even number of photons (most often, two photons), 
while o-Ps undergoes annihilation with emission of an odd number of photons (most often, three photons). 
Therefore, due to the smallness of the fine-structure coupling constant and to the differences 
in phase space volume available for decays into different number of photons, 
the average lifetimes of o-Ps and p-Ps differ by more than three orders of magnitude. 
The lifetime of p-Ps in a vacuum is equal to $\tau_{p-Ps}$~$\approx$~0.125~ns~\cite{al-ramadhan}, 
whereas the average lifetime of o-Ps amounts to $\tau_{o-Ps}$~$\approx$~142~ns~\cite{vallery,jinnouchi,m10}. 
Such large difference in lifetimes enables a very efficient experimental disentangling of these states.

\section{Motivation to test discrete symmetries in decays of positronium atoms}
\subsection{Tests of charge conjugation symmetry}
Violation of C symmetry in gravitational, strong and electromagnetic interaction 
was not observed so far.  
The best limit in systems of quarks was set for $\pi^0 \to 3\gamma$ decay 
and amounts to 3.1$\times$10$^{-8}$ at 90\%~CL~\cite{m14,m15}. 
In the framework of the J-PET experiment we intend to study the C symmetry 
in the leptonic system searching 
for the C-forbidden decays of the positronium atoms. 
Positronium, as a system of fermions must be anti-symmetric under exchange of its components. 
Thus, as mentioned in the previous section, ground state of ortho-positronium as symmetric in space (L~=~0) and in spin (S~=~1) 
must be C-symmetry odd, 
and para-positronium as symmetric in space and anti-symmetric in spin (S~=~0) must be C-symmetry even.
Photons are C-odd and thus the C eigenvalue of $n$ photons is equal to $(-1)^n$. 
Therefore, C symmetry forbids decays of ortho-positronium into even number of photons
and para-positronium into an odd number of photons.
It is important to stress that such decays cannot occur due to the possible mixing 
between various positronium states because there are no different positronium states 
with the same J$^{P}$ and opposite C parity~\cite{m10}.
The branching ratio for these decays 
(calculated based on the Standard Model) is of the order of 10$^{-10}$ - 10$^{-9}$~\cite{m13,m16}. 
Hence, observation of the branching ratio for e.g.  
$o\!-\!Ps \to 4\gamma$ or $p\!-\!Ps \to 3\gamma$ 
significantly greater than expected on the basis of the Standard Model 
would suggest violation of the C symmetry by the electromagnetic interactions. 
So far we know only upper limits of these branching ratios amounting to:

BR($o\!-\!Ps \to 4\gamma / o\!-\!Ps \to 3\gamma) < 2.6 \times 10^{-6}$ at 90\%~CL~\cite{m17},

BR($p\!-\!Ps \to 3\gamma / p\!-\!Ps \to 2\gamma) < 2.8 \times 10^{-6}$ at 68\%~CL~\cite{m18}, 

BR($p\!-\!Ps \to 5\gamma / p\!-\!Ps \to 2\gamma) < 2.7 \times 10^{-7}$ at 90\%~CL~\cite{m19},\\
where the best limit~\cite{m19} 
was achieved in the Lawrence Berkeley National Laboratory 
in USA by means of the Gammasphere detector, 
a spectrometer for nuclear structure research~\cite{m19}. 
These values are still more than two orders of magnitude larger than expected 
for the C-conserving process which would imitate C-violation due to the vacuum polarization effects.

\subsection{Tests of CP symmetry}
Violation of the CP symmetry has been observed for the first time in 1964 
in the decays of K$_L$ mesons~\cite{m20}. 
Interestingly the CP violation in K$_S$ mesons has not been observed so far. 
The best limit was set by the KLOE-2 collaboration in 2013~\cite{m21,m22}. 
K mesons remained the only systems for which such violation was observed 
for almost forty years. First indications of the 
CP violation in the B mesons decays were reported only in the year 2001 by the Belle Collaboration 
at KEK in Japan~\cite{m23} and by the BABAR Collaboration at Stanford Linear Accelerator 
Center SLAC in USA~\cite{m24}. 
In 1973 M. Kobayashi and T. Maskawa~\cite{m25} explained the CP violation 
in quark systems via the presence of the complex phase in the quark transition matrix, 
extending the model of N. Cabibbo~\cite{m26}, and suggesting the existence of three quark generations 
(at those days only u, d and s quarks were known but indeed c, b, and t quarks were discovered later). 
For this achievement Kobayashi and Maskawa were awarded the Nobel prize in 2008.  
Andrei Sakharov has argued that violations of C and CP symmetries 
are necessary requirements for the explanation 
of the observed excess of matter over anti-matter~\cite{m1}. 
However, the presently known sources of the CP symmetry violations 
are still by far too small and can account for only about 10$^{-9}$ fraction 
of the observed excess of matter over anti-matter~\cite{m2}. 
Therefore, many particle physics experiments as e.g. LHCb experiment~\cite{m5} 
at CERN or \mbox{Belle-II} experiment~\cite{m6} at KEK
plan to search for CP symmetry violation effects in hadrons with 
upgraded detectors and beam intensities. 

Though the violation of the CP symmetry is small 
it has been shown in several contributions 
that is has a crucial impact on various domains in physics. 
For instance, the role of time is still puzzling and controversial; 
whereas in Quantum Mechanics position is well described in theory by an operator, 
time usually is treated as a parameter. 
Time operator models exist, in particular a certain one denoted the temporal wave function model 
that extends the Born rule to the time domain~\cite{beatrix1}. 
It had been in no conflict with any known experiment; 
indeed, the falsification of the model turned out to be 
due to the violation of the CP symmetry~\cite{beatrix1},
though tiny. 
Another example is given by showing that correlations stronger than those allowed 
by classical physics are related to the violation of the CP symmetry~\cite{beatrix2,beatrix3}
and thus, in principle, providing the security for quantum cryptographic protocols~\cite{beatrix4}.

The search for CP violation was conducted also in other systems, 
as e.g. in the $\eta \to \pi^+\pi^-e^+e^-$ decay~\cite{DanWASA} 
which does not change the flavor quantum number. 
Such CP symmetry breaking is not included in the Standard Model. 
This CP non-conservation would manifest itself in the asymmetry 
in the angular distribution between the emission planes of $e^+e^-$ and $\pi^+\pi^-$ pairs. 
So far such a violation was not observed and the most accurate 
results were obtained with the KLOE experiment~\cite{m28}.
Concurrently, other experiments may test the CP-symmetry 
violation in processes involving purely leptonic systems. 
To this end in USA neutrinos from Fermilab are sent to NOvA detector~\cite{m29} 
810 km away, and in Japan the T2K collaboration~\cite{m30} 
sends neutrinos from J-PARC in Tokai to 
295~km away Super-Kamiokande detector in Kamioka. 
The most sensitive signature of the CP-symmetry violation which is expected  
from these experiments is a difference between the probability of muon neutrino oscillations 
into electron neutrino  P($\nu_\mu \to \nu_e$) 
and the probability of its CP symmetric process which is oscillations 
of muon anti-neutrino into electron anti-neutrino P($\bar{\nu}_\mu \to \bar{\nu}_e$). 
However, it is estimated~\cite{m7,m8}  that NOvA and T2K experiments can provide a CP violation 
discovery potential up to 2$\sigma$ only in the year 2020 and up to 3$\sigma$ in 2024. 
Thus, the study of the CP symmetry violation in lepton systems 
is one of the most exciting challenges of today’s particle physics.

We intend to use the J-PET detector to search for CP symmetry violation in decays of positronium atoms.  
We plan to measure the expectation value of the CP-odd operators constructed 
from the momentum and polarization vectors of photons and polarization of positronium atoms. 
Examples of such operators are given hereafter in Table 1. 
Observation of the non-zero expectation value of the CP-odd operator would imply the CP non-invariance. 
There is a very strong limitation on the electric dipole moment of electron 
$|d_e| < 10.5 \times 10^{-28}$~e~cm at 90\%~CL~\cite{m14,m32}. 
It puts limitations of the order of $10^{-18}$~\cite{m33} 
on CP violation driven by the Standard Model mechanisms 
such as e.g. one-photon exchange $ee\gamma$ or $ee2\gamma$ and $ee3\gamma$ processes.  
However, it does not preclude a possible much larger violation of CP symmetry 
in the positronium annihilation due to exotic non Standard Model mechanisms~\cite{m34}. 
On the other hand the photon-photon interaction in the final state due to
the vacuum polarization may mimic CP non-invariance on the level of 10$^{-9}$~\cite{m13,m16}, 
and effects due to the weak interaction can lead to a violation at the order of 10$^{-14}$~\cite{m10}. 
The present best upper limit on the CP violation in the decays of ortho-positronium 
atoms was determined at the University of Tokyo and amounts to 4.9$\times$10$^{-3}$ at 90\%~CL~\cite{m35}. 
Thus, there is still more than six orders of magnitude difference between the present 
experimental upper limit and the CP violation expected due to the photon-photon interaction. 

\subsection{Tests of T symmetry (time reversal invariance)}
Time reversal symmetry (T) is epistemologically extremely attractive.
However, although one can reverse the direction of the motion in space, 
one cannot reverse the direction of the elapsing time.  
Therefore, in order to study the time reversal violation 
one investigates expectation values for the T-odd operators 
of the non-degenerate stationary states~\cite{m36} 
as e.g. electric dipole moment  (EDM) of the particles with spin, 
or one compares processes which are related 
by exchange of initial and final states equivalent to reversing the time.
In the latter case one reverses only a motion, 
and therefore the time reversal invariance is often referred to as 
“motion reversal” symmetry~\cite{m36}. 
For such studies, the anti-unitary character of the T operator makes the experimental studies 
of the time reversal invariance more challenging than other symmetries~\cite{m36},
since it requires abilities of preparing the initial and final states of the process 
in a fully controlled way.

Only recently in 2012, in the experiment conducted by the BABAR collaboration 
with the use of the SLAC linear accelerator at Stanford University, 
the time reversal violation was observed 
independently of the CP or CPT symmetries~\cite{m12}. 
Based on the quantum entangled B mesons produced in the decays of the Y(4S) meson 
and using the ideas described in ref.~\cite{m37}, the BABAR group has shown that 
the probabilities of the time reversal symmetric transitions between the 
flavor eigenstates ($B_0,\bar{B_0}$) and CP eigenstates ($B_+,B_-$) are different. 
For example, it was shown that the probability of the transition between 
flavor $\bar{B^0}$ and CP~=~-1 states ($\bar{B^0} \to B_-$) 
is different from the probability for its T symmetric  process:  $B_- \to \bar{B^0}$.  
Importantly, T, CP and CPT symmetries applied to the $\bar{B^0} \to  B_-$ 
transition lead to the different processes:  
T($\bar{B^0} \to B_-)~=~B_- \to \bar{B^0}$;  
CP($\bar{B^0} \to  B_-)~=~ B^0 \to B_-$; 
CPT($\bar{B^0} \to B_-)~=~ B_- \to B^0$.
The obtained result is consistent with the level of CP symmetry violation 
and the assumption of the CPT conservation. 

So far none of the experiments reported limits 
on the T symmetry violations in the decays of positronium. 
The known final state interactions of photons are expected 
to mimic the T violation at the level of 10$^{-9}$~\cite{m13,m16}.
All of the previous investigations with positronium, 
which tested the discrete symmetries odd operators, 
were based on the products of photons momenta and positronium spin vectors. 
We plan to extend the study to other operators,
taking advantages of properties of the J-PET tomography scanner, 
which enables to determine the polarization of photons (see section~\ref{photo-polarization}).
Therefore, using the J-PET detector the T symmetry 
can be investigated by searching for the possible non-zero expectation 
values of the T-odd operators which can be constructed 
from momentum and polarization directions of photons 
originating from the decays of the o-Ps atoms as well as from the spin of these atoms (see section~5).

It is important to stress, that so far all experiments searching for the non-zero expectation value 
of the T-odd operators provided no signals of T-symmetry violation 
and hence the K and B mesons remains the only systems 
for which the time reversal symmetry violation was observed. 

\subsection{Tests of CPT symmetry}
CPT symmetry being combination of the above described discrete symmetries 
of charge conjugation C, spatial parity P and reversal in time T, 
constitutes a fundamental law of quantum field theory 
resulting from the locality and unitarity of interactions 
and from the Lorentz invariance~\cite{m38,m39}. 
There are known processes breaking invariance of C, P and T separately 
and even few processes breaking combined CP symmetry, 
but so far violation of CPT invariance has not been observed. 
The breaking of this symmetry would imply violation of one 
or more of the above mentioned fundamental assumptions, 
and at the same time, as proven by Greenberg, 
it would imply the breaking of the Lorentz symmetry~\cite{m40}. 
We cannot a priori exclude that unitarity, 
locality or Lorentz invariance are not broken in the regime of very high energies, 
where the quantum gravity effects may play a significant role~\cite{m41}. 
There exist many unverified models and theories among which as an example 
it is worth to mention the extensions of the standard model non-invariant 
with respect to CPT and Lorentz symmetry as proposed by Kostelecky~\cite{m42}. 
However, it is not possible for the moment to predict precisely how such effects 
could manifest themselves in phenomena involving particles 
and anti-particles at energies available in laboratories.
Searching for CPT symmetry breaking in processes involving 
various kinds of particles should be treated as complementary. 
Therefore, in many physics laboratories CPT invariance is tested 
in various physical systems,  
for example via comparison of meson and its anti-meson properties~\cite{m43}, 
or via comparison of properties of hydrogen and anti-hydrogen~\cite{m44,m45,marcin1,marcin2}. 
One of the most sensitive tests were made~\cite{m46} 
and are planned~\cite{m28} based on the quantum interferometry 
of entangled neutral kaons~\cite{Antonio-T}.

It is worth noting that with respect to the quark sector, 
relatively poor experimental limits of CPT symmetry violation 
were achieved in processes involving decays of positronium atoms. 
The best limit so far on the CPT violation in positronium decays 
was set on the level of 0.3\% and it was obtained using the Gammasphere 
array of Compton suppressed high-purity germanium detectors~\cite{m47}. 
Thus, this is still six order of magnitude larger than the possible contribution 
from radiative corrections which may mimic CPT violation at the level of 10$^{-9}$~\cite{m16}. 
Analogously, as in the case of the above discussed CP and T symmetries, 
the tests of CPT invariance will be conducted using the J-PET detector 
via the measurement of the expectation values of the CPT-odd operators listed in Table~1. 

\section{J-PET: Jagiellonian Positron Emission Tomograph}
\begin{figure}[htb]
\centerline{%
\includegraphics[width=7.0cm]{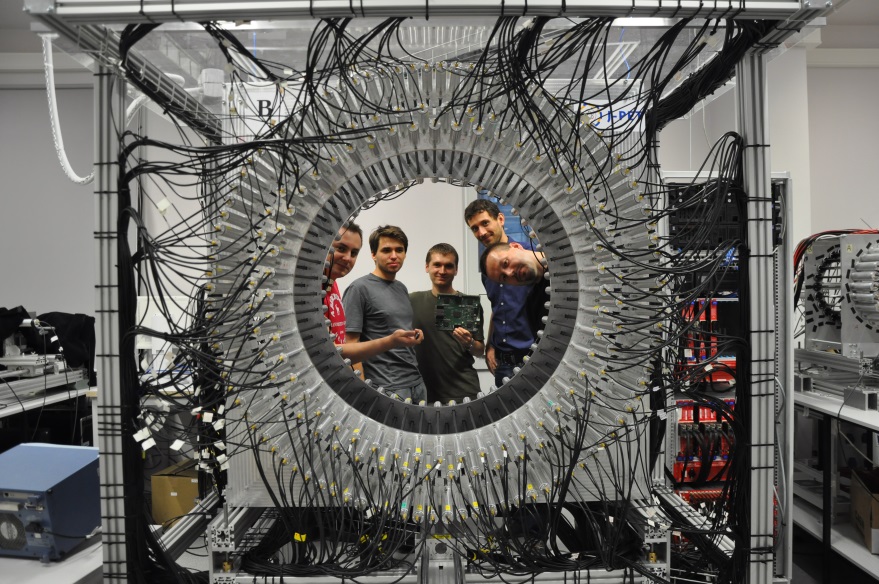}
\includegraphics[width=4.5cm]{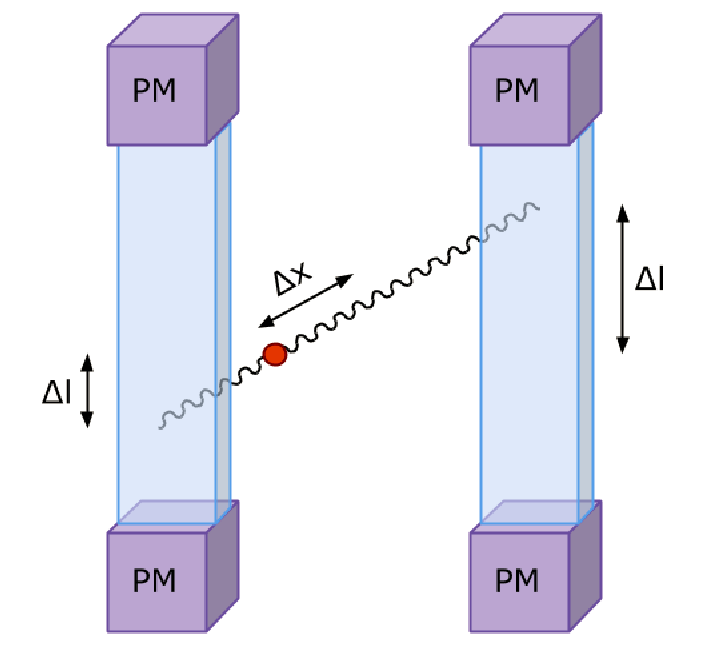}}
\caption{
(Left) Photo of the J-PET detector with a few members of the J-PET group. 
The active inner part of the detector has a cylindrical shape with the length of 50~cm 
and diameter of 85~cm. 
The J-PET tomography scanner is made of three layers of plastic scintillator strips 
wrapped with the Vikuiti specular foil covered with an additional light-tight foil~(black). 
The scintillators are optically connected at two ends to Hamamatsu R9800 vacuum tube photomultipliers (gray). 
(Right) Schematic view of the two detection modules~\cite{J6}. 
A single detection module consists of a EJ-230 scintillator strip read out by two photomultipliers 
labeled PM. In the first approximation the hit distance from the center of the scintillator ($\Delta L$) 
is determined based on time difference measured at both ends of the scintillator strip.  
In case of the two-photon annihilation, the position ($\Delta x$) along the “line of annihilation” 
is determined from time difference measured between two modules. 
In practice, more advanced methods of hit-time and hit-position 
were developed which take advantage of the variation of the signal 
shape as a function of the hit-position~\cite{J4,J5,J6}.  In the case of three photon annihilation,
the positronium decay point is reconstructed based on the trilateration method~\cite{AlekNIM}.
\label{Fig1-JPET}
}
\end{figure}

The J-PET scanner is built out of strips of organic scintillator,
forming a cylinder (left panel of Fig.~\ref{Fig1-JPET})~\cite{mpatent1,mpatent2}. 
Light signals from each strip are converted to electrical signals by photomultipliers 
placed at opposite ends of the strip~\cite{J1}. 
The position and time of reaction of photons in the detector material 
is determined based on the arrival time of light signals 
to the ends of the scintillator strips (right panel of Fig.~\ref{Fig1-JPET}). 
The signals are probed in the voltage domain with the accuracy 
of about 30~ps by a newly developed multi-threshold digital electronics~\cite{J2,palkapatent} 
and the data are collected by the novel trigger-less 
and reconfigurable data acquisition system~\cite{J3,korcylpatent,GrzegorzActa}.
The readout data is streamed to the Central Controller Module 
and then, further, to permanent storage~\cite{GrzegorzActa}. For the data processing 
and simulations a  dedicated software framework was developed~\cite{WojciechActaA,WojciechNukleonika,WojciechActaB}.
The hit-position and hit-time of photons in the scintillator strips
are reconstructed by the dedicated reconstruction 
methods based on the compressive sensing theory~\cite{J4,J5} 
and the library of synchronized model signals~\cite{J6,Neha,PMActaA}. 
The hit-time and hit-position reconstruction procedures are further developed, 
but it is important to note that at present the achieved resolution 
for the determination of the hit-time of the annihilation 
photons is about 0.1~ns for energy deposition about 0.27~MeV~\cite{J1,J6}. 

\subsection{Positronium production}
\begin{figure}[htb]
\centerline{%
\includegraphics[width=1.0\textwidth]{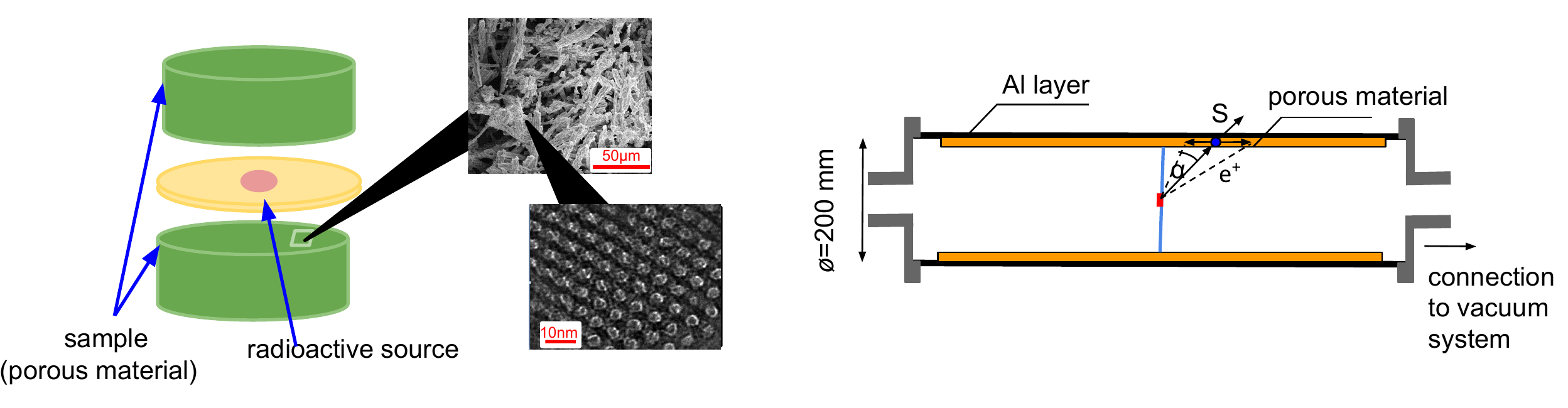} 
}
\caption{
(Left)
Pictorial presentation of the positronium source 
with the radioactive isotope surrounded by the porous material. 
Inserted figures were taken 
for the SBA-15 porous material using SEM (upper figure) and TEM (lower one) techniques. 
Bars indicate 50~$\mu$m~(upper figure) and 10~nm (lower figure).
The positronium source will be kept in the annihilation chamber
with the vacuum sufficient to suppress effectively ortho-para 
conversion with paramagnetic molecules of air.
(Right)
Schematic view of a spin-linear polarized positronium target 
in the form of cylinder surrounding the positron source.
Longitudinally polarized positrons are emitted from the beta-plus 
radioactive source placed in the center of the cylinder.
Ortho-positronium atoms with spin $\vec{S}$ are formed 
in the cylindrical layer of the porous material.
The direction of the polarization will be determined with the uncertainty of 
$\pm \alpha$ determined by the precision of the o-Ps annihilation point reconstruction
using a trilateration-based method introduced recently in the reference~\cite{AlekNIM}. 
\label{target}
}
\end{figure}

Positronia will be produced by positrons interacting with electrons in the porous materials.
Three different kinds of positronium targets are planned to be prepared, which will be optimized
for the production of (i) unpolarized positronium atoms, (ii) linearly spin-polarized positronium atoms,
and (iii) tensor spin-polarized positronium atoms. 
In each case a beta-plus radioactive isotope of $^{22}$Na is planned to be used as positrons source.
However, for some tests of systematic effects we consider also the use 
of $^{68}$Ge and $^{44}$Sc~\cite{Jastrzebski} isotopes.
For the production of unpolarized and spin-tensor polarized positronium
a radioisotope $^{22}$Na affixed in 7.5~$\mu$m Kapton foil will be used. 
The source will be inserted between two layers of target material (sandwich configuration),
as it is depicted schematically in the left panel of Fig.~\ref{target}.
%
The positronium produced in the material may disintegrate not only via internal annihilation,
but also due to the interaction with electrons from surrounding molecules.
Therefore, in order to minimize systematic uncertainties 
in studies of the symmetries violation in the ortho-positronium decays 
the life-time of ortho-positronium and its production probability in the material
should be maximized~\cite{BozenaActa}. 
More specifically the fraction ($f_{3\gamma}$) of $o\!\!-\!\!Ps \to 3\gamma$ decay rate to the overall annihilation rate,
as well as the fraction of events from ortho-positronium decays to all the 3$\gamma$ annihilations 
($f_{3\gamma}(o\!-\!\!Ps) = N(o\!-\!\!Ps \to 3\gamma) / N(3\gamma)$) should be as large as possible.
The best known materials for such experiments 
are silica aerogels~\cite{Bozena25,Bozena26} 
in which empty space constitutes more than 90\% of the total volume~\cite{Bozena28}.
Recently, we have performed studies~\cite{BozenaActa}
by measuring the life-time and production probability of ortho-positronium atoms, 
as well as the 3$\gamma$ fraction of their annihilation
in samples of commercial aerogels (IC3100, IC3110, IC3120 and LA1000), 
Amberlite porous polymer XAD4 (CAS 37380-42-0)
and silica porous material SBA-15 synthesised in the Faculty of Chemistry 
of Maria Curie-Sklodowska University~\cite{Bozena29}. 
We found that the life-time is the largest for the aerogel IC3100 and amounts to about 132~ns,
and that the fractions $f_{3\gamma}$ and $f_{3\gamma}(o\!-\!\!Ps)$ 
are the largest for the XAD-4 porous polymer.
In particular, we observed that for this polymer the fraction $f_{3\gamma}$ 
varies between 24.4\% to 28.9\%
depending on the measurement method, and that $f_{3\gamma}(o\!-\!\!Ps)$ is as large as 99.7\%~\cite{BozenaActa}.

\subsection{Linear polarization of ortho-positronium atoms}
\begin{figure}[htb]
\centerline{%
\includegraphics[width=6.4cm]{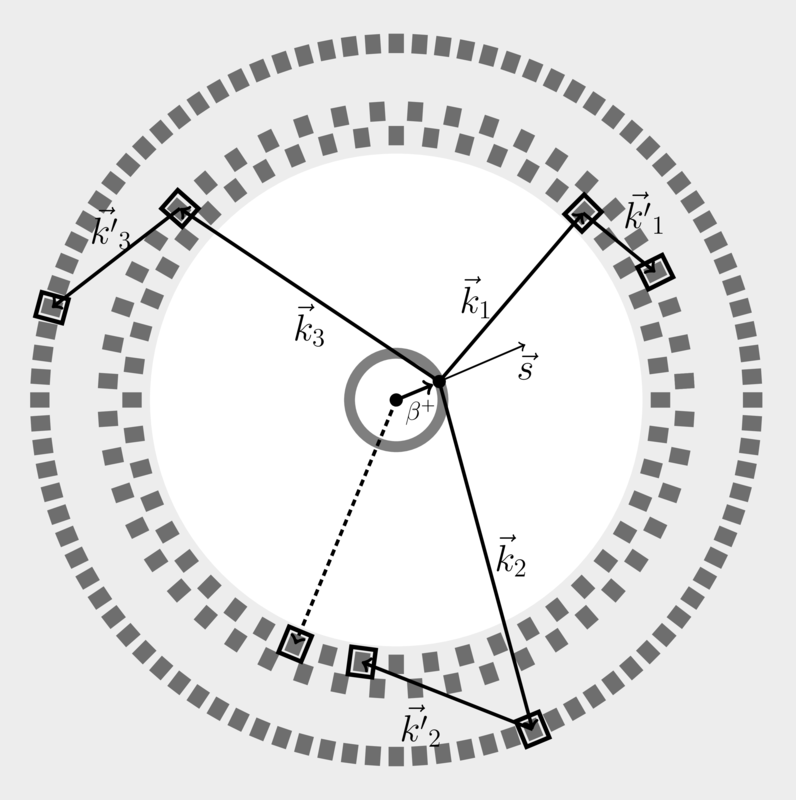} \hspace{-0.4cm} \includegraphics[width=6.4cm]{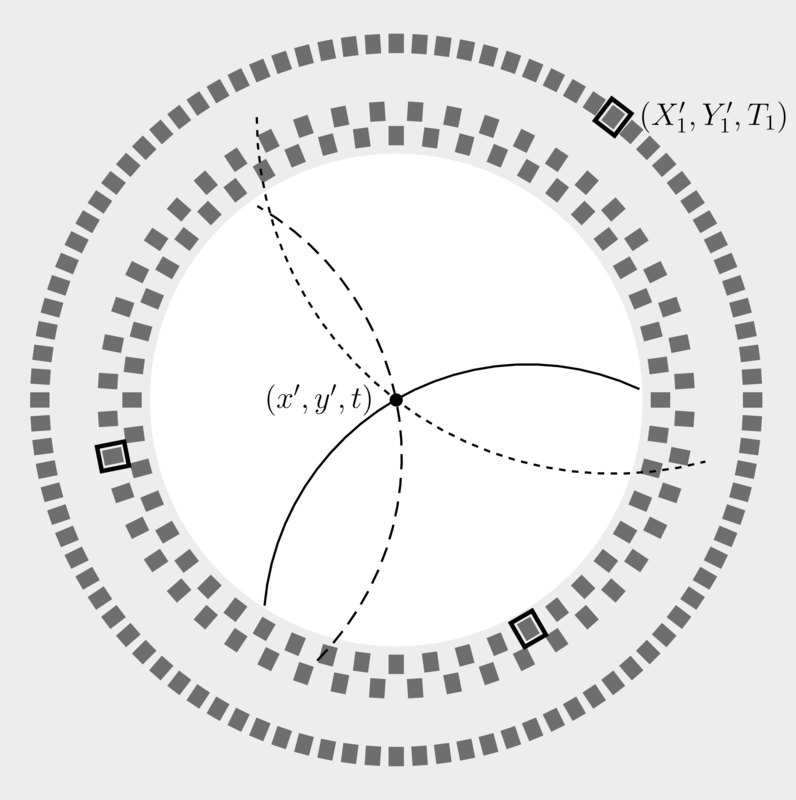}
}
\caption{
(Left)
Schematic view of the cross section of the J-PET detector (rectangles) 
with the linearly polarized positronium source indicated as a circle surrounding the center. 
A projection onto the plane perpendicular to the detector axis is shown. 
Source may be realized as an aerogel cylinder (see Fig.~\ref{target})
whose diameter, housing material and thickness still need to be optimized.
Superimposed arrows 
indicate momentum and spin vectors involved in the typical event 
of ortho-positronium production with spin ($\vec{S}$).
A black dot in the center indicates a sodium $^{22}$Na radioactive source emitting 
longitudinally polarized positron ($\vec{\beta^+}$) from  parity violating beta-decay 
and prompt gamma quantum (dotted arrow) from the deexcitation of daughter nucleus $^{22}$Ne.
Positronium (dot marked in the cylinder) decays into three photons ($\vec{k_1},\vec{k_2},\vec{k_3}$).
Each photon may interact via Compton effect in the scintillator strips (marked by rectangular rims),
and the secondary photons ($\vec{k_1^\prime},\vec{k_2^\prime},\vec{k_3^\prime}$) may also react in the 
scintillator strips.
(Right)
Scheme of the reconstruction of ortho-positronium annihilation time and point ($x^\prime,y^\prime,t$)
inside the decay plane. 
Hit-positions (rectangular rim surrounding the hit scintillator strip) for each recorded photon constitute 
the center of a circle 
(e.g. $X_1^\prime, Y_1^\prime,T_1$) 
describing possible photon origin points 
where radius of the circle depends on the difference between the hit-time and the annihilation time $t$. 
The decay point is found as an intersection of the three circles.
The right figure is adapted from reference~\cite{AlekNIM}.
\label{j-pet-polarization}
}
\end{figure}

The geometry and properties of the J-PET detector enable to design the positronium 
source such that the vector polarization of produced ortho-positronium can be determined.  
Due to the parity violation in the beta-decay the emitted positrons 
are longitudinally polarized with the polarization vector equal to $\vec{P}~=~\vec{v}/c$,
where $\vec{v}$ denotes the positron velocity.  This effect was used e.g. in the Gammasphere experiments
where the target was built from a hemisphere of silicon dioxide aerogel 
with the radioactive source in the center of the sphere~\cite{m47}. 
For the J-PET experiment we plan to construct a spin-polarized positronium source as shown 
in the right panel of Fig.~\ref{target} and Fig.~\ref{j-pet-polarization}.

The positron emitted from the source placed in the center of the detector interacts in the 
cylindrical layer of the porous material.  
The created positronium annihilates into three photons
(see Fig.~\ref{j-pet-polarization}) and the time and position of their interactions in the scintillator strips
enables to reconstruct the position of annihilation by means of the trilateration method. 
The trilateration-based reconstruction 
of ortho-positronium annihilation position and time 
was recently introduced in references~\cite{m50,AlekNIM}. 
Detailed simulations taking into account the properties of the J-PET detector
indicate that this method enables to reconstruct the annihilation position with 
the spatial resolution of about 1~cm to 2~cm~\cite{AlekNIM}, 
and the annihilation time with the resolution of about 0.1~ns~\cite{AlekNIM}.
In the Global Positioning System (GPS) a trilateration is based on 
measurements of time and position of signals by four satellites.
In the case of the reconstruction of ortho-positronium location, the lack of a fourth reference
system is compensated by the momentum conservation which implies 
that the decay point of ortho-positronium ($o\!-\!\!Ps \to 3\gamma$)
and the momentum vectors of photons are contained in a single plane. 
The trilateration reconstruction method is illustrated 
in the right panel of Fig.~\ref{j-pet-polarization}.
Knowing the emission and annihilation position of the positron 
one can reconstruct the direction of motion of the positron and, hence, 
the direction of the spin for the positronium.

In the case of the $^{22}$Na source, the average degree of the positronium polarization 
is equal to P~=~$<\!\!v\!\!>$/c~=~0.67, and for $^{68}$Ge polarization P amounts to 0.9~\cite{m47,Coleman}.
This polarization is to large extent preserved during the thermalization process~\cite{m48}.
Its losses in the material are estimated 
to be about 17\% for positrons from $^{68}$Ge and 8\% for positrons from $^{22}$Na~\cite{m49}.  
Taking into account that only 2/3 of created ortho-positronium atoms possess 
spins parallel to the spins of the positrons~\cite{m16}
it was estimated that the average polarization of ortho-positronium 
(achievable by irradiating the porous materials with positrons from the beta-plus isotope) 
is equal to about 0.41 when using $^{22}$Na and 0.50 when applying $^{68}$Ge~\cite{m47,m49}.  
For positrons emitted in the cone with an opening angle of 2$\alpha$  
the average polarization is decreased by a factor of $(1~+~\cos(\alpha))/2$~\cite{Coleman}. 
Therefore, in practice, the polarization degree is further lowered due to the finite angular 
resolution of the determination of the positrons direction.

With 
the cylindrical target 
and 
the J-PET detector 
shown schematically in the right panel of Fig.~\ref{target}
and the right panel of Fig.~\ref{j-pet-polarization}, respectively,
we can achieve an angular resolution 
for the positron direction determination of about 15$^\circ$~\cite{AlekNIM}. 
Therefore, the loss of polarization due to the uncertainty of the 
determination of positron direction will amount to about 2\% only.
In the case of the Gammasphere experiment the annihilation points were not reconstructed, 
so only a polarization averaged over the total hemisphere was used, 
thus leading to an effective polarization loss by a factor of two.  

\subsection{Tensor spin-polarized source of positronium atoms}
Tensor polarization of o-Ps atoms is defined as: 
\begin{equation}
   P_2~=~(N_{+1} - 2N_0 + N_{-1}) / (N_{+1} + N_0 + N_{-1}),  
\end{equation}
where N$_{+1}$, N$_0$, and N$_{-1}$ denote the number of atoms with spin projection along 
the quantization axis equal to +1, 0 and -1, respectively.  
The quantization axis will be provided by the direction of 
the external magnetic field.
Each of the spin projection can occur with the same probability. 
A non-zero value of tensor polarization may be effectively generated 
by the off-line selection of events corresponding to decays 
of ortho-positronium in the static magnetic field.
The method makes use of the fact that 
in the magnetic field the o-Ps state with m~=~0 mixes with p-Ps. 
This mixing changes significantly the lifetime of o-Ps in m~=~0 state. 
E.g. in the magnetic field of ~5~kG, in a aerogel the o-Ps(m~=~0) 
life-time decreases from about 126~ns to about 22~ns~\cite{m35}, 
whereas the life-time of o-Ps states with m~=~+1 or m~$=-1$   
is not affected by the magnetic field. 
Therefore, by choosing a  life-time interval of o-Ps one can vary the ratio of 
N$_0$ to (N$_{+1}$ + N$_{-1}$) and, hence, vary the polarization P$_2$. 
By using a magnetic field of $\sim$5~kG and a time window from 50 to 130~ns (in aerogel) 
a polarization degree of P$_2$~=~0.87 was achieved in the experiment described 
in the reference~\cite{m35}.  
The design of a proper system for the generation of the static 
magnetic field is one of the tasks being presently realized by the J-PET collaboration. 
The first preliminary studies by means of the Vizimag simulation software 
indicate that using a method of active shielding allows to generate a field  
in the order of 1~T at the source in the center of the detector with the fringe 
field less than the Earth's magnetic field at the photomultipliers 
(distant by about 60~cm). Problems with photomultipliers in the magnetic field
can be reduced by using a SiPM readout~\cite{PMB}.

\subsection{Monitoring of ortho-positronium spin-polarization}
In the case of the tensor polarization the quantization axis is defined by the 
direction of the magnetic field, and in the case of the linear polarization we can 
determine it on the event by event basis as a direction of the positron motion. 
The degree of the tensor polarization may be monitored based on the measurement of the angular
distribution between the normal to the decay plane and the quantization axis, where the decay plane
is defined as the plane containing momenta of photons from the $o\!-\!Ps \to 3\gamma$ decay.
The distribution of the angle $\theta_S$ between the spin direction of o-Ps  
and the normal to the decay plane is proportional to the term (1~+~$\cos{^2\theta_S}$) for m~=~0, 
and to the term $\frac{1}{2}$(3~-~$\cos^2\theta_S$) for m~=~$\pm$~1~\cite{Bernreuther1981}. 
Therefore, we can determine the degree of the tensor polarization 
by the comparison of the angular distribution of the normal to the decay plane 
with respect to the direction of the magnetic field for the cases when magnetic field is switched on and off.  
Without magnetic field an isotropic distribution is expected and with the magnetic 
field the polarization degree will be derived from the amplitude of the term proportional to $\cos^2\theta_S$. 
In the extreme cases if P$_2$~=~-2 the expected distribution should be proportional to (1~+~$\cos{^2\theta_S}$), 
whereas for P$_2$~=~1 it should behave as (3~-~$\cos^2\theta_S$).
We can also control how the degree of  the tensor polarization changes 
with the life-time interval of o-Ps chosen in the analysis.

The angular distribution of $\theta_S$ can be also used to control the systematic uncertainties 
in the case of the studies with the linearly spin-polarized \mbox{o-Ps}.
For example, by the comparison of the angular distributions when using different positron sources 
(e.g. such as $^{22}$Na and $^{68}$Ge) as it was done in the case of the Gammashere experiment~\cite{m47}.

\subsection{Polarisation of photons}
\label{photo-polarization}
\begin{figure}[htb]
\centerline{%
\includegraphics[width=1.0\textwidth]{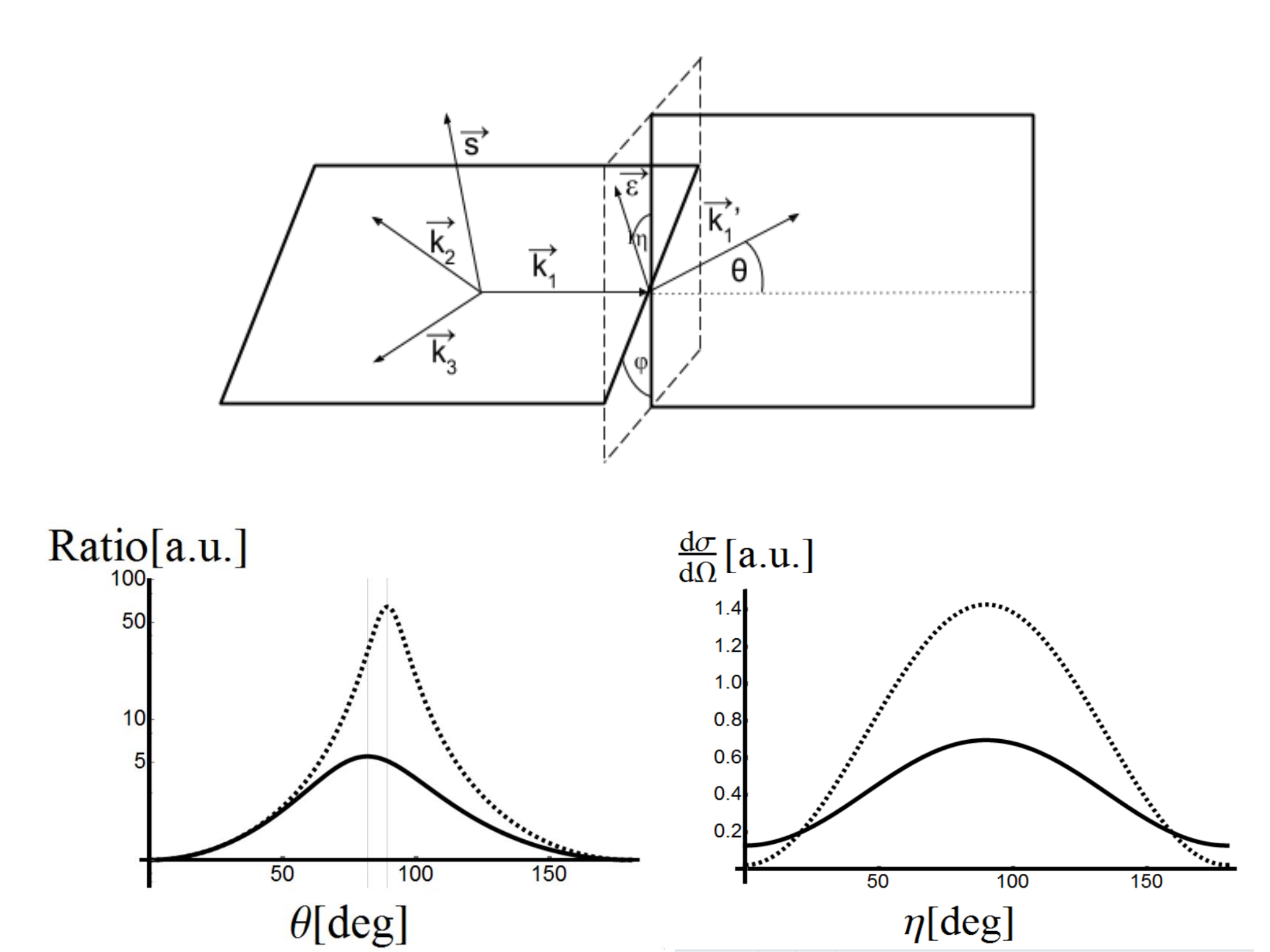}
}
\caption{
(Upper panel) Schematic illustration of the decay plane of ortho-positronium 
and the scattering plane of one of the photons. 
$\vec{k_i}$ denotes momentum vector of i-th photon, 
$\vec{\epsilon}$ denotes the electric vector of photon, 
$\theta$ stands for the scattering angle, 
$\phi$ indicates angle between the decay plane of ortho-positronium and the scattering plane of the photon denoted by $\vec{k_1}$. 
$\eta$ denotes the angle between the scattering plane and the electric vector of photon with momentum $\vec{k_1}$, 
and 
$\vec{S}$ indicates the spin of ortho-positronium. 
(Lower left) A ratio  of [$d\sigma/d\Omega(\eta=90^{\circ})]/[d\sigma/d\Omega(\eta=0^{\circ})$] as a function of the scattering angle $\theta$. 
Solid line represents result for k$_1$~=~511~keV, and dashed line for k$_1$~=~100~keV.
(Lower right) $d\sigma/d\Omega$ in arbitrary units shown 
        as a function of the $\eta$ angle for the $\theta$ fixed to 82$^\circ$ in case of the 511~keV photon (solid line)
        and for $\theta$ fixed to 89$^\circ$ in case of the 100~keV photon (dashed line).
\label{plaszczyzny}
}
\end{figure}

J-PET scanner is built from plastic scintillators strips. 
Therefore, photons from the positronium decay are interacting via Compton effect and 
a fraction of them may undergo two or more scatterings 
in different strips (see Fig.~\ref{j-pet-polarization} and reference~\cite{J8}). 
Since gamma quantum is a transverse electromagnetic wave, 
and since Compton scattering is at most likely in the plane perpendicular 
to the electric vector of the photon~\cite{KleinNishina,Evans}, 
we can determine the direction of its linear polarization $\vec{\epsilon_i}$ 
e.g. by constructing $\vec{\epsilon_i} = \vec{k_i} \times \vec{k^\prime_i}$,
where $\vec{k_i}$ and $\vec{k^\prime_i}$ denote momentum vectors 
of i-th gamma quantum before and after the Compton scattering, respectively (as indicated in Fig.\ref{j-pet-polarization}).
The ortho-positronium decay plane and the scattering 
plane for one of the photons are shown schematically in Fig.~\ref{plaszczyzny}.
$\theta$ denotes the scattering angle between the directions of propagation 
of the primary photon $\vec{k}$ and the scattered photon $\vec{k^\prime}$.  
It is important to note that, independently of the value of $\theta$,
the probability of the scattering has its maximum value 
when the scattering plane is perpendicular to the direction 
of the electric vector of the primary photon ($\eta$~=~90$^\circ$). 
The corresponding cross section reads~\cite{KleinNishina,Evans}:
\begin{equation}
d\sigma/d\Omega \sim (k^\prime/k)^2 (k/k^\prime + k^\prime/k - 2\sin^2\!{\theta} \, cos^2{\eta}), 
\end{equation}
where the meaning of the angles is explained in Fig.~\ref{plaszczyzny}.
The above formulae indicates that this probability is maximum for ($\eta$~=~90$^\circ$), 
but it is not necessarily vanishing when the scattering plane is parallel to $\vec{\epsilon}$.
Lower left panel of Fig.~\ref{plaszczyzny} indicates the ratio  
of [$d\sigma/d\Omega(\eta=90^{\circ})$]/[$d\sigma/d\Omega(\eta=0^{\circ})$] as a function of the scattering angle $\theta$.
Energy of photons from the $o\!-\!Ps \to 3\gamma$ decays ranges from 0 to 511~keV.
In this energy range the ratio changes significantly. For example for the 511~keV photons
it is maximal at $\theta$ of about 82$^\circ$ 
and for photons with energy of 100~keV the maximum is at about 89$^\circ$. 
The dependences of $d\sigma/d\Omega$ on the $\eta$ angle
are shown in the right panel of Fig.~\ref{plaszczyzny}.
This dependence of the cross sections sets the limit for the achievable
resolution for the determination of the direction
of the $\vec{\epsilon}$ with the ansatz that $\vec{\epsilon_i} = \vec{k_i} \times \vec{k^\prime_i}$.
At present a quantitative estimation is ongoing.
However, it is worth to stress that, independently of the determination of the photons polarization,
the ability of the J-PET detector to determine the angle $\phi$ between the decay and scattering planes,
as well as the angles between o-Ps spin and the scattering planes of photons
opens possibilities for experimental definition of orthogonal states of photon
and, hence, enables studies of the of multi-partite entanglement of high energy photons
originating from the positronium annihilation~\cite{Beatrix}.

\subsection{Application of the PALS method with the J-PET detector}
\begin{figure}[htb]
\centerline{%
\includegraphics[width=6.5cm]{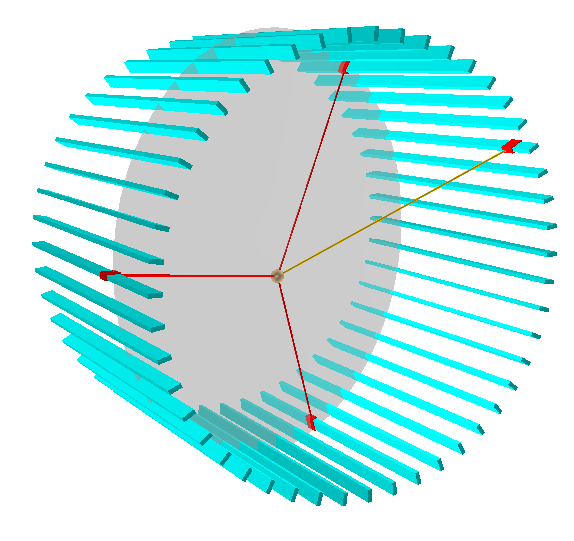}
\includegraphics[width=6.0cm]{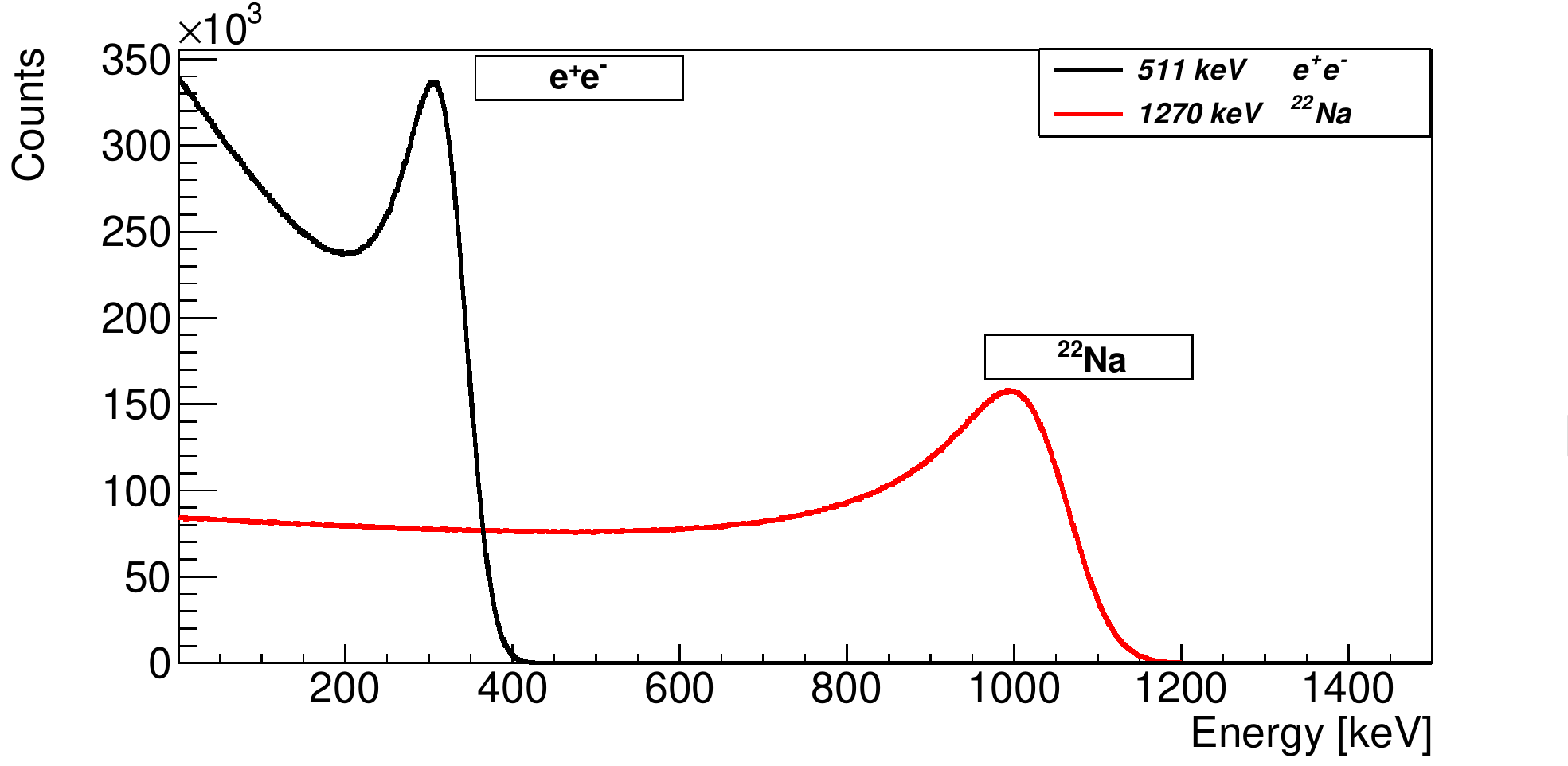}
}
\caption{
(Left)
Schematic view of the inner layer of the J-PET detector with superimposed decay plane of 
$o\!-\!Ps \to  3\gamma$ process. 
Lines inside the plane indicate annihilation photons and the  
line pointing out of the plane indicates a deexcitation photon used for the start time determination.  
(Right) 
Simulated energy loss spectra for annihilation (0.511~MeV) 
and deexcitation photons from the $^{22}$Na decay (1.27~MeV). 
The shown distributions include the energy resolution of the J-PET detector~\cite{J1}.
\label{PALS}
}
\end{figure}

The J-PET detector enables to determine the life-time of produced positronium atoms on the event-by-event basis.
For this purpose, not only  photons from the $e^+e^-$ annihilation 
but also the deexcitation photon 
from the excited daughter nucleus originating from the beta-plus 
decay will be registered. This will permit to measure differences between 
the time of formation and the time of annihilation of positronium atoms. 
These times can be reconstructed since with the J-PET detector we can reconstruct 
hit-time and hit-position of the photons, as well as the annihilation position~\cite{AlekNIM,m50,m51}. 
In the case of unpolarized source, 
for the time of formation of the positronium atom we will use the reconstructed time 
of the emission of the deexcitation photon.
This approximation leads to negligible uncertainties, since the time between the
emission of the positron and the formation of the positronium, 
as well as the average life-time of the $^{22}$Ne$^*$ excited nucleus 
($^{22}Na \to ^{22}\!\!Ne^* \, e^+ \, \nu_e \to ^{22}\!\!Ne  \, \gamma \, e^+ \,  \nu_e$),
are in the order of few~ps. 
Due to the momentum conservation the three photons from the o-Ps decay  
will move inside a plane comprising the annihilation point (Fig.~\ref{PALS}(left)).
However, the direction of movement of the deexcitation quantum, 
not being correlated with the annihilation photons, 
will be distributed isotropically with respect to the decay plane. 
The annihilation and deexcitation photons can be disentangled 
based on the combination of at least three criteria:
(i) the measured energy deposition, 
(ii) the distance between the decay plane and the annihilation place, 
and 
(iii) the hit-time difference which is fixed for a given annihilation 
and hit positions for annihilation photons. 
Fig.~\ref{PALS}(right) shows energy loss spectrum expected for the highest energy 
photons from the positronium annihilation compared to the spectra 
expected from the deexcitation photons from $^{22}$Na isotope. 
The results were obtained taking into account the experimental energy resolution of the J-PET detector~\cite{J1}. 

\subsection{Simulations of $o\!-\!Ps \to 3\gamma$ decays} 
Simulation-based studies are the subject of the ongoing research of the J-PET collaboration. 
However, here we present preliminary results which show that the competing 
accuracy with respect to the previous experiments can be achieved for the discussed studies.  
Simulations are conducted using the GEANT-4 Application for Tomographic Emission (GATE)~\cite{J8} 
supplemented with our own simulation procedures which are necessary 
in order to account for annihilation processes to more than $2\gamma$, 
because such processes are not included in the standard commonly 
applied simulation packages. 
The J-PET detector is based on the plastic scintillators and, in practice, due to the low atomic number 
of this material, photons in the energy range between  50~keV and 511~keV 
interact predominantly via the Compton effect.  
We have simulated the response of the detector to $o\!-\!Ps \to 3\gamma$ process taking into account: 
(i) the energy dependence of the $o\!-\!Ps \to 3\gamma$ transition amplitude according to the predictions 
based on the quantum electrodynamics~\cite{m54}, 
(ii) the energy dependence of the total cross section for the Compton scattering, 
(iii) the energy distribution of the scattered electrons, 
(iv) the geometry of the J-PET system as well as its energy resolution~\cite{J1}. 
The correctness of the simulation procedures was corroborated by the comparison 
between measured and simulated distributions for 511~keV photons~\cite{J6}.
The details of the simulations will be described in the forthcoming article~\cite{Daria-ops}.

\subsection{Reduction of the instrumental background}
\begin{figure}[htb]
\centerline{%
\includegraphics[width=6.5cm]{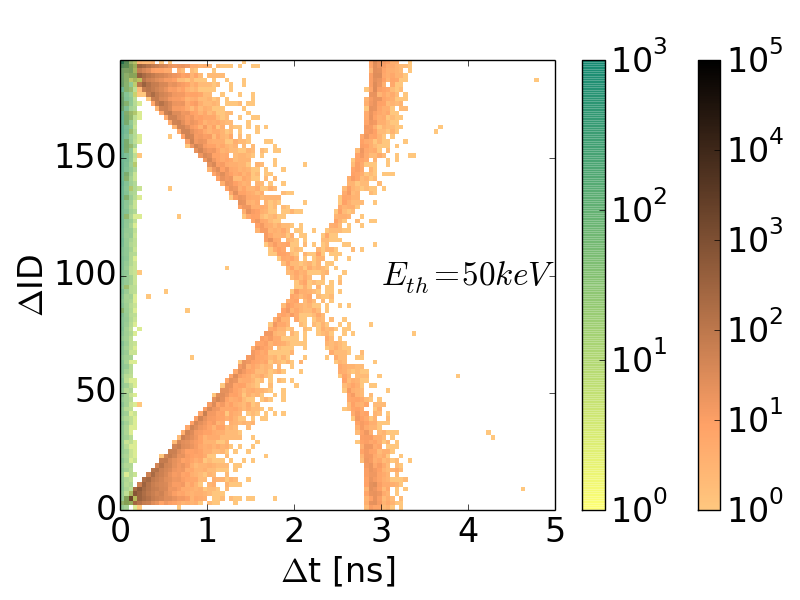}
\includegraphics[width=6.5cm]{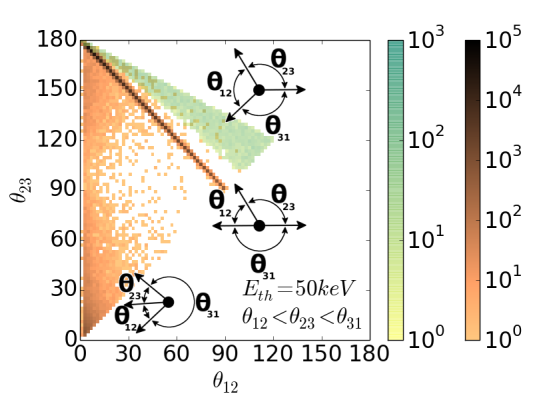}
}
\caption{
(Left) Simulated distributions of differences between detectors ID ($\Delta ID$) 
and differences of hit-times ($\Delta t$) for events with three hits registered 
from the annihilation $e^+e^- \to 2\gamma$ (points concentrated in diagonal and anti-diagonal bands) 
and $o\!-\!Ps \to 3\gamma$ (points concentrated in vertical band close to $\Delta t=0$).  
The module ID increases monotonically with the growth of the azimuthal angle between the vertical 
direction and the line connecting the center of the J-PET detector and a given module~\cite{J8}. 
(Right) Distribution of relative angles between reconstructed directions of photons. 
The numbering of photons was assigned such that $\theta_{12} < \theta_{23} < \theta_{31}$. 
Shown distributions were obtained requiring three hits each with energy deposition larger than E$_{th}$~=~50~keV.  
Points below and at diagonal band show results for simulations of $e^+e^- \to 2\gamma$ 
and 
points above the diagonal band correspond
to $o-Ps \to 3\gamma$. Typical topology of $o\!-\!Ps \to 3\gamma$ 
and two kinds of background events are indicated. 
The figures are adapted from~\cite{J8,Daria}.
\label{dt-vs-id}
}
\end{figure}

The J-PET detector was designed for the medical imaging purposes
and it allows for the identification of the  $e^+e^- \to  2\gamma$ events. 
In addition, the very good angular resolution of about one degree 
and time resolution of about 100~ps of the J-PET detector allows for the significant suppression 
of the background also for the studies of $o\!-\!\!Ps \to 3\gamma$ events.  
Here we show only rough estimations 
in order to demonstrate the background suppression capability of the J-PET system.
The most dangerous instrumental background is due to the secondary scattering 
of photons originating from the  $2\gamma$ events which may mimic the registration of $3\gamma$.  
However, we can disentangle true and false $3\gamma$ events  based on 
(i) the relation between  position of the individual detectors (ID) 
    and the time difference between registered hits ($\Delta t$),  
(ii) the angular correlation of angles between the direction of photons,  
and 
(iii) the distance between the origin of the annihilation (position of the annihilation chamber) 
and the decay plane.   
In Fig.~\ref{dt-vs-id}, we show two example spectra  with the expected $\Delta ID$ vs. $\Delta t$  
and $\theta_{23}$~vs.~$\theta_{12}$ distributions for $o\!-\!\!Ps \to 3\gamma$ events 
with the superimposed distributions originating from to the $e^+e^-\to 2\gamma$ 
events which resulted in three hits in the detector. 
For the true $o\!-\!\!Ps \to 3\gamma$ events  the $\Delta t$ between different hits 
in the detector should be equal to zero, 
and,  therefore,  the true and false $3\gamma$ events only slightly overlap. 
Similarly, after ordering the relative angles ($\theta_{12} < \theta_{23} < \theta_{31}$)  
the true and false events have very small (almost none) overlap region 
at the $\theta_{23}$~vs.~$\theta_{12}$ correlation plot.  
Applying relatively conservative criteria on all of these mentioned observables 
(e.g.  $\theta_{23} > 185 - \theta_{12}$  and  $\Delta t < 0.3$~ns at Fig.~\ref{dt-vs-id})
we expect to reduce the background from  2$\gamma$ annihilation by a factor larger  than $10^9$ 
(including detection acceptance and efficiency). 
This high reduction  of  background is crucial because the efficiency 
for the simultaneous detection of 3$\gamma$ from the o-Ps decay is  small. 

\subsection{Physical background due to the positronium production method}
Due to the short life-time of p-Ps (~0.125~ns) and short average time of direct annihilations (about~0.5~ns) 
their misidentification as $o\!-\!Ps\to 3\gamma$  events can be reduced to a negligible level 
by the requirement that the time difference between the deexcitation and annihilation photons
is larger than e.g. 50~ns as used e.g. in reference~\cite{m35}. 
However, such condition does not eliminate the background contributions originating 
from interactions of o-Ps with electrons at the surfaces of voids in the porous material.  
The rate of this kind of o-Ps loss may be estimated from its life-time in the material 
used for the target, and  e.g. in the 
XAD-4 porous polymer~\cite{BozenaActa}  
it is equal to (1~-~$\tau_{aero}/\tau_{vacuum}$)~=~(1~-~91/142)$~\approx$~0.36, 
and for the aerogel IC3100 it amounts to about 0.07.
Thus, about 7\% to 36\% of o-Ps may undergo pick-off annihilations 
or ortho-para spin conversion (due to the spin-orbit interaction or due to electron exchange). 
Further on we will assume that each of these processes contributes by 20\%, 
however it is important to stress that this is  a very conservative assumption since e.g. 
conversion processes are strongly suppressed in the  aerogels in vacuum as it will be in our case. 

In priciple, the sample of events recognized as originating from 
o-Ps decay ($N_{o-Ps}$) would include events from the following processes: 
true decays of $o\!-\!Ps \to 3\gamma$ ($N_{o-Ps} \to 3\gamma$); 
pick-off with subsequent prompt annihilation to $3\gamma$ ($N_{3\gamma-pick-off}$); 
pick-off with subsequent prompt annihilation to $2\gamma$ 
misidentified as $3\gamma$ due to secondary scatterings ($N_{2\gamma-pick-off}$); 
conversion of o-Ps to p-Ps with subsequent C symmetry violating decay to $3\gamma$ ($N_{3\gamma-conv}$); 
and conversion of o-Ps to p-Ps with subsequent annihilation to $2\gamma$ misidentified as $3\gamma$ 
due to the secondary scatterings ($N_{2\gamma-conv}$). 
Taking into account the upper limits for the C symmetry violations, the detection and reconstruction efficiencies,
as well as the background suppression discussed above, we found that the largest contribution 
(by almost two orders of magnitude larger than the others)
which may mimic the symmetry violation originates from the 
pick-off process with subsequent prompt annihilation to $3\gamma$ ($N_{3\gamma-pick-off}$). 
The conservative upper limit of this background contribution may be estimated as:
$N_{3\gamma-pick-off}$  / $N_{o-Ps}  < 0.2/370$~$\approx$~5$\times$10$^{-4}$.

The experiment described in this article aims at the determination 
of the expectation value of correlation operators 
which are odd with respect to the studied symmetries. 
The nonzero values of such operators would reflect 
at the non-zero asymmetry of counts defined appropriately 
for each studied symmetry. 
It is a matter of further investigations to estimate a possible asymmetry 
due to the 
background
discussed 
above. 
However, assuming arbitrarily but rather conservatively 
that the asymmetry of $3\gamma\!-pick\!-\!off$  background is less than 10\% 
we would obtain the systematic uncertainty of about 5$\times$10$^{-5}$ 
which is by two orders of magnitude lower than the present experimental limits~\cite{m35,m47}. 
Moreover, it should be stressed that the measurement of the true $2\gamma$ 
events with high statistics will allow for the precise control of the amount of this background contribution. 
In addition, the possible contribution to this asymmetry from the true $o\!-\!Ps \to 3\gamma$ 
decays and the background events can be disentangled by the comparison of results obtained  
for materials with different voids sizes. The asymmetry originating from the background contributions 
changes with the void size, whereas the contribution from the true $o\!-\!Ps \to 3\gamma$ decay remains constant. 
Therefore, we expect to reach a systematic sensitivity below 5$\times$10$^{-5}$.

\section{Observables for the discrete symmetries}
We intend to test discrete symmetries by the determination 
of the expectation value of the operators listed in Table~\ref{table1}. 
Observation of the non-zero expectation value of any of these operators 
would imply a non-invariance of these symmetries 
for which the given operator is odd-symmetric (marked in the table with $"-"$).
For example, the CPT symmetry may be tested by measuring the angular correlations between 
the spin of the ortho-positronium atom and its decay plane. 
CPT symmetry breaking may manifest itself as an asymmetry of the $\vec{S} \cdot (\vec{k_1} \times \vec{k_2})$ operator. 
Its non-zero expectation value signals a difference in the probability between cases
when spin vector $\vec{S}$ of o-Ps atom is pointing up and down with respect to the normal 
to the decay plane defined as $\vec{k_1} \times \vec{k_2}$. 
Thus CPT symmetry violation would manifest itself as an asymmetry in the decay plane orientation 
with respect to the initial spin of the o-Ps atom. 
The decay plane may be oriented by the ordering of photons
according to the descending momentum: $k_1 > k_2 > k_3$. 
\begin{table}[h!]
\centering
\caption{
Operators for the $o\!-\!\!Ps\to3\gamma$ process 
and their properties with respect to the C, P, T, CP and CPT symmetries.  
New operators (including $\vec{\epsilon}$) available at J-PET are shown in the last three rows.
\label{table1}
}
\vspace{0.5cm}
\begin{tabular}{|c|c|c|c|c|c|}
\hline
 \textbf{Operator} & \textbf{C} & \textbf{P} & \textbf{T} & \textbf{CP} & \textbf{CPT} \\\hline
 $\vec{S}   \cdot \vec{k_1}$                                                & $+$ & $-$ & $+$ & $-$ & $-$   \\\hline
 $\vec{S}   \cdot (\vec{k_1} \times \vec{k_2}$)                              & $+$ & $+$ & $-$ & $+$ & $-$   \\\hline
 $(\vec{S}  \cdot \vec{k_1}) (\vec{S} \cdot (\vec{k_1} \times \vec{k_2}))$ & $+$ & $-$ & $-$ & $-$ & $+$   \\\hline
 $\vec{k_2} \times \vec{\epsilon_1}$                                         & $+$ & $-$ & $-$ & $-$ & $+$   \\\hline
 $\vec{S}   \cdot  \vec{\epsilon_1}$                                         & $+$ & $+$ & $-$ & $+$ & $-$   \\\hline
 $\vec{S}   \cdot (\vec{k_2} \times \vec{\epsilon_1})$                      & $+$ & $-$ & $+$ & $-$ & $-$   \\\hline
\end{tabular}
\end{table}
J-PET detector does not allow for the direct measurement of the photon energy. 
However, we can reconstruct the momentum of each photon from the $o\!-\!Ps\to 3\gamma$ process
based on the measurement 
of the hit-positions and annihilation point which allows for the determination 
of the relative angles between the photons. 
The reconstruction of these angles and the use of energy and momentum conservations 
permit to reconstruct the momenta of all decay photons. 
As preparatory studies we have performed Monte-Carlo simulations 
of the response of the J-PET detector to the $o\!-\!Ps \to 3\gamma$ decay 
and reconstructed the photons energies and propagation angles. 
Taking into account the experimentally determined position and time resolutions 
of the detector~\cite{J1,J6} 
we have obtained angular resolution of about 1 degree and energy resolution of about 5 keV~\cite{Daria-ops}. 

In order to test the C symmetry we will search for the C violating decays of ortho-positronium 
($o\!-\!Ps \to 4\gamma$) and para-positronium ($p\!-\!Ps \to 3\gamma$). 
However, to reach a higher sensitivity in the studies of the C symmetry 
violation not only the number of photons will be taken into account, 
but we will also make the comparison between symmetric and asymmetric configurations 
of photons. We will take advantage of the fact that para-positronium 
decay $p\!-\!Ps\to 3\gamma$  in the symmetric configuration 
(with photons emitted with directions towards the vertices of the equilateral triangle) 
is forbidden not only by the C-symmetry but also by the Bose-Einstein statistics 
and rotational symmetry (angular momentum conservation)~\cite{m10}.  
Therefore, the C symmetry violation in the $p\!-\!Ps \to 3\gamma$ decay may also manifest itself 
as a variation of the ratio of $3\gamma$ decays in the symmetric and non-symmetric configurations, 
while the ratio of ortho-positronium to para-positronium is varied. 
The latter may be achieved by the variation of the gas pressure 
in the positronium production targets or by the variation 
of the static magnetic field inside the target. 
The time resolution of the J-PET tomograph of about~0.1~ns~\cite{J1} 
will be of great advantage in selecting smaller time window corresponding 
to the production of para-positronium and hence in reducing the background 
from the C-allowed decays of ortho-positronium $o\!-\!Ps\to 3\gamma$. 
Analogously, it is favorable to study the C-symmetry violation via the comparison 
of the $o\!-\!Ps \to  4\gamma$ decay rate into symmetric and non-symmetric configurations. 
In this case in the symmetric configuration photons are emitted along 
the directions defined by the vertices of the tetrahedron. 
The C-allowed decay $p\!-\!Ps \to 4\gamma$ into symmetric configuration 
is forbidden via rotational symmetry~\cite{m52}. 
Therefore, the physical background due to the C-allowed $o\!-\!Ps \to 4\gamma$ decay 
is negligible in this symmetric configuration and thus only an instrumental 
background will limit the accuracy. 
In addition, in the tetrahedral configuration, a C symmetry violation in $o\!-\!Ps \to 4\gamma$
decay is expected to be maximal~\cite{m52,m53}.

\section{Summary}
The Jagiellonian Positron Emission Tomograph (J-PET)
was constructed as a prototype of a cost-effective scanner for the 
simultaneous metabolic imaging of the whole human body~\cite{J1,J6,PMB}.
It is optimized for the detection 
of photons from the electron-positron annihilation
with high time- and angular-resolutions,
thus providing new opportunities 
for research with photons originating from the decays of positronium atoms
in fundamental physics, as well as in life and material sciences~\cite{P3,EwelinaNukleonika,AnnaActa,AnnaNukleonika}.

The C, CP, T and CPT symmetries are of fundamental importance in physics. 
Violation of T or CP invariance in purely leptonic systems have never been seen so far. 
Based on known mechanisms of C and CP violations,  
one cannot explain the large asymmetry between matter and antimatter in the observable Universe. 
The above facts constitute motivation for the elaboration of the experimental proposal
presented in this article which comprises
studies of asymmetries 
between leptons and antileptons at low energies with the J-PET detector.
In this article we described the potential of the J-PET detector for tests 
of C, CP, T and CPT symmetries in the decays of para- and ortho-positronium atoms. 
The tests will be based on the determination of the expectation values of the discrete-symmetries-odd operators 
which may be constructed from the momentum and polarization vectors of photons 
and from the spin vector of ortho-positronium atom.

With respect to the previous experiments perforemd with crystal based detectors,
J-PET built from plastic scintillators, provides 
superior time resolution, higher grantularity, lower pile-ups,
and opportunity of determining photon's polarization. 
These features allow us to expect improvement
by more than an order of magnitude in tests of discrete symmetries 
in decays of positronium atoms.

\section{Acknowledgements}
We acknowledge valuable discussions with Dr Jan Wawryszczuk and technical and administrative support by  
A.~Heczko, M.~Kajetanowicz, W.~Migda\l, 
and the financial support by The Polish National Center for Research and Development
through grants INNOTECH-K1/IN1/64/159174/NCBR/12 and LIDER-274/L-6/14/NCBR/2015,
The Foundation for Polish Science through MPD program and the EU, MSHE Grant No. POIG.02.03.00-161 00-013/09,
and The Marian Smoluchowski Krakow Research Consortium ”Matter-Energy-Future”.
B.C.H. gratefully acknowledges the Austrian Science Fund FWF-P26783 and FWF-23627.


\begin{thebibliography}{}
\bibitem{m1} A. D. Sakharov, \textit{Pisma Zh. Eksp. Teor. Fiz.} \textbf{5}, 32 (1967).
\bibitem{m2} D. N. Spergel \textit{et al., Astrophys. J. Suppl.} \textbf{148}, 175 (2003) \newline [arXiv:astro-ph/0302209].
\bibitem{m3} M. Fukugita, T. Yanagida, \textit{Phys. Lett. B} \textbf{174}, 45 (1986). 
\bibitem{m4} W. Buchmuller, T. Yanagita, \textit{Ann. Rev. Nucl. Part. Sci.} \textbf{55}, 311 (2005) [arXiv:hep-ph/0502169].
\bibitem{m5} R. Aaij \textit{et al., Eur. Phys. J.} \textbf{C73}, 2373 (2013) [arXiv:1208.3355 [hep-ex]].
\bibitem{m6} I. Adachi \textit{et al., JINST} \textbf{9}, C07017 (2014). 
\bibitem{m7} M. Ghosh \textit{et al., Nucl. Phys. B} \textbf{884}, 274 (2014) [arXiv:1401.7243 [hep-ph]].
\bibitem{m8} M. Ghosh \textit{et al., Phys. Rev. D} \textbf{89}, 011301(R) (2014) [arXiv:1306.2500 [hep-ph]]. 
\bibitem{m9} M. Deutsch, \textit{Phys. Rev.} \textbf{82}, 455 (1951). (Nobel Prize in 1956).
\bibitem{al-ramadhan} A. H. Al-Ramadhan, D. W. Gidley, \textit{Phys. Rev. Lett.} \textbf{72},  1632 (1994).
\bibitem{vallery} R. S. Vallery, P. W. Zitzewitz, D. W. Gidley, \textit{Phys. Rev. Lett.} \textbf{90},  203402 (2003).
\bibitem{jinnouchi} O. Jinnouchi, S. Asai, T. Kobayashi, \textit{Phys. Lett.} \textbf{B 572},  117 (3003).
\bibitem{m10} M. S. Sozzi, Discrete Symmetries and CP Violation. From Experiment to Theory, Oxford University Press (2008).
\bibitem{badertscher} A. Badertscher et al. (2007). "An Improved Limit on Invisible Decays of
\bibitem{m11} A. Angelopoulos \textit{et al., Phys. Lett. B} \textbf{444}, 43 (1998). 
\bibitem{m12} J. P. Lees \textit{et al., Phys. Rev. Lett.} \textbf{109}, 211801 (2012) \newline [arXiv:1207.5832 [hep-ex]].
\bibitem{m20} J. H. Christenson, J. W. Cronin, V. L. Fitch, R. Turlay, \textit{Phys. Rev. Lett.} \textbf{13}, 138 (1964).
\bibitem{m13} W. Bernreuther \textit{et al., Z. Phys. C} \textbf{41}, 143 (1988).
\bibitem{m14} K.A. Olive \textit{et al.} (Particle Data Group), \textit{Chin. Phys. C} \textbf{38}, 090001 (2014). 
\bibitem{m15} J. McDonough \textit{et al., Phys. Rev. D} \textbf{38}, 2121 (1988).
\bibitem{m16} B. K. Arbic \textit{et al., Phys. Rev. A} \textbf{37}, 3189 (1988). 
\bibitem{m17} J. Yang \textit{et al., Phys. Rev. A} \textbf{54}, 1952 (1996).
\bibitem{m18} A. P. Mills, S. Berko, \textit{Phys. Rev. Lett.}, \textbf{18}, 420 (1967). 
\bibitem{m19} P. A.Vetter, S. J. Freedman, \textit{Phys. Rev. A} \textbf{66}, 052505 (2002). 
\bibitem{m21} D. Babusci \textit{et al., Phys. Lett. B} \textbf{723}, 54 (2013) [arXiv:1301.7623 [hep-ex]]. 
\bibitem{m22} M. Silarski, PhD thesis, Jagiellonian University (2012) [arXiv:1302.4427].
\bibitem{m23} A. Abashian \textit{et al., Phys. Rev. Lett.} \textbf{86}, 2509 (2001) [arXiv:hep-ex/0102018].
\bibitem{m24} B. Aubert \textit{et al., Phys. Rev. Lett.} \textbf{86}, 2515 (2001) [arXiv:hep-ex/0102030].
\bibitem{m25} M. Kobayashi, T. Maskawa, \textit{Prog. Theor. Phys.} \textbf{49}, 652 (1973). (Nobel prize in 2008)
\bibitem{m26} N. Cabibbo, \textit{Phys. Rev. Lett.} \textbf{10}, 531 (1963).
\bibitem{beatrix1} T. Durt, A. Di Domenico, B. Hiesmayr, [arXiv:1512.08437]
\bibitem{beatrix2} B. C. Hiesmayr \textit{et al., Eur. Phys. J} \textbf{C 72}, 1856 (2012)
\bibitem{beatrix3} B. C. Hiesmayr, \textit{Found. Phys. Lett.} \textbf{14}, 231 (2001).
\bibitem{beatrix4} B. C. Hiesmayr, \textit{J. Phys. Conf. Ser.}  \textbf{631}, 012067 (2015).
\bibitem{DanWASA} P. Adlarson \textit{et al.,} (2015) [arXiv:1509.06588].
\bibitem{m28} G. Amelino-Camelia \textit{et al., Eur. Phys. J. C} \textbf{68}, 619 (2010) \newline [arXiv:1003.3868 [hep-ex]].
\bibitem{m29} D. S. Ayres \textit{et al.,} [arXiv:hep-ex/0503053]; J. Bian, [arXiv:1309.7898]. 
\bibitem{m30} K. Abe \textit{et al., Phys. Rev. Lett.} \textbf{112}, 061802 (2014) [arXiv:1311.4750 [hep-ex]].
\bibitem{m32} C. A. Baker \textit{et al., Phys. Rev. Lett.} \textbf{97}, 131801 (2006) [arXiv:hep-ex/0602020]. 
\bibitem{m33} H. Y. Cheng, \textit{Phys. Rev. D} \textbf{28}, 150 (1983).
\bibitem{m34} S.N. Gninenko, N.V. Krasnikov, A. Rubbia, \textit{Mod. Phys. Lett. A} \textbf{17}, 1713 (2002).
\bibitem{m35} T. Yamazaki \textit{et al., Phys. Rev. Lett.} \textbf{104}, 083401 (2010) \newline [arXiv:0912.0843 [hep-ex]].
\bibitem{PMB} P. Moskal \textit{et al.,} [arXive:1602.02058]; \textit{Phys. Med. Biol.} (2016) in print.
\bibitem{beatrix} T. Durt, A. Di Domenico, B. Hiesmayr, arXiv:1512.0843.
\bibitem{m36} J. Bernabeu, F. Martinez-Vidal, \textit{Rev. Mod. Phys.} \textbf{87}, 165 (2015) [arXiv:1410.1742 [hep-ph]].
\bibitem{m37} J. Bernabeu \textit{et al., JHEP} \textbf{1208}, 64 (2012) [arXiv:1203.0171 [hep-ph]].
\bibitem{m38} J. S. Bell, \textit{Proc. R. Soc. London Ser. A} \textbf{231}, 479 (1955).
\bibitem{m39} G. L{\"u}ders, \textit{Ann. Phys.} \textbf{2}, 1 (1957); G. L{\"u}ders, \textit{Ann. Phys. (N.Y.)} \textbf{281}, 1004 (2000).
\bibitem{m40} O. W. Greenberg, \textit{Phys. Rev. Lett.} \textbf{89}, 231602 (2002) [arXiv:hep-ph/0201258]. 
\bibitem{m41} S. Hawking, \textit{Commun. Math. Phys.} \textbf{87}, 395 (1982).
\bibitem{m42} V. A. Kostelecky, N. Russell, \textit{Rev. Mod. Phys.} \textbf{83}, 11 (2011) \newline [arXiv:0801.0287 [hep-ph]].
\bibitem{m43} A. Di Domenico, \textit{Frascati Phys. Ser.} \textbf{43}, (2007).
\bibitem{m44} D. F. Phillips \textit{et al., Phys. Rev. D} \textbf{63}, 111101 (2001) \newline [arXiv:physics/0008230 [physics.atom-ph]].
\bibitem{m45} G. Gabrielse \textit{et al., Phys. Rev. Lett.} \textbf{82}, 3198 (1999).
\bibitem{marcin1} G.~Gabrielse \textit{et al., Phys. Rev. Lett.} \textbf{108}, 113002 (2012).
\bibitem{marcin2} J. DiSciacca \textit{et al., Phys. Rev. Lett.} \textbf{110}, 130801 (2013).
\bibitem{m46} D. Babusci \textit{et al., Phys. Lett. B} \textbf{730}, 89 (2014) [arXiv:1312.6818 [hep-ex]].
\bibitem{Antonio-T} J. Bernabeu, A. Di Domenico, P. Villanueva-Perez, \textit{Nucl. Phys. B} \textbf{868}, 102 (2013) [arXiv:1208.0773 [hep-ph]].
\bibitem{m47} P. A. Vetter, S. J. Freedman, \textit{Phys. Rev. Lett.} \textbf{91}, 263401 (2003).
\bibitem{mpatent2} P. Moskal, Patent number: WO2011008118-A2; PL388556-A1; US2012175523-A1; EP2454611-A2; JP2012533733-W.
\bibitem{palkapatent} M. Pa\l ka, P. Moskal, Patent number: WO2015028600-A1.
\bibitem{korcylpatent} G. Korcyl, P. Moskal, M. Kajetanowicz, M. Pa\l ka, Patent number: WO2015028594-A1.
\bibitem{J1} P. Moskal \textit{et al., Nucl. Instrum. and Meth.} {\bf A 764}, 317 (2014) \newline [arXiv:1407.7395 [physics.ins-det]].
\bibitem{J2} M. Pa{\l}ka \textit{et al., Bio-Algorithms and Med-Systems} \textbf{10}, 41 (2014) [arXiv:1311.6127 [physics.ins-det]].
\bibitem{GrzegorzActa} G. Korcyl \textit{et al., Acta Phys. Pol.} \textbf{47}, (2016), this proceedings.
\bibitem{WojciechNukleonika} W. Krzemie\'n  \textit{et al., Nukleonika} \textbf{60}, 745 (2015) [arXiv:1508.02751].
\bibitem{WojciechActaA} W. Krzemie\'n \textit{et al., Acta Phys. Polon. A} \textbf{127}, 1491 (2015) \newline [arXiv:1503.00465 [physics.ins-det]].
\bibitem{WojciechActaB} W. Krzemie\'n \textit{et al.,  Acta Phys. Pol. B} \textbf{47} (2016), these proceedings.
\bibitem{J3} G. Korcyl \textit{et al., Bio-Algorithms and Med-Systems} \textbf{10}, 37 (2014).
\bibitem{J4} L. Raczy\'nski \textit{et al., Nucl. Instrum. and Meth. A} \textbf{764}, 186 (2014) [arXiv:1407.8293 [physics.ins-det]].  
\bibitem{J5} L. Raczy\'nski \textit{et al., Nucl. Instrum. and Meth. A} \textbf{786}, 105 (2015) [arXiv:1503.05188 [physics.ins-det]].
\bibitem{J6} P. Moskal \textit{et al., Nucl. Inst. and Meth. A} \textbf{775}, 54 (2015) [arXiv:1412.6963].
\bibitem{AlekNIM} A. Gajos \textit{et al.}, Patent number: PCT/PL2015/050038; Nucl. Instrum. and Meth. submitted.
\bibitem{Neha} N. Sharma \textit{et al., Nukleonika} \textbf{60}, 765 (2015) \newline [arXiv:1508.07463 [physics.ins-det]].
\bibitem{PMActaA} P. Moskal \textit{et al., Acta Phys. Polon. A} \textbf{127}, 1495 (2015) \newline [arXiv:1502.07886 [physics.ins-det]].
\bibitem{J8} P. Kowalski \textit{et al., Acta Phys. Polon. A} \textbf{127}, 1505 (2015) \newline [arXiv:1502.04532 [physics.ins-det]].
\bibitem{J9} P. Moskal \textit{et al., Phys. Med. Biol.}, in print.
\bibitem{Jastrzebski} K. Szkliniarz \textit{et al., Acta Phys. Pol. A} \textbf{127}, 1471 (2015). 
\bibitem{m48} J. Van House, P. W. Zitzewitz, \textit{Phys. Rev. A} \textbf{29}, 96 (1984);
              P. W. Zitzewitz \textit{et al., Phys. Rev. Lett.} \textbf{43}, 1281 (1979).
\bibitem{m49} J. Yang \textit{et al., Jpn. J. Appl. Phys.} \textbf{36}, 3764 (1997).  
\bibitem{m50} A. Gajos, Diploma thesis, Jagiellonian University, (2013).
\bibitem{m51} A. Gajos, \textit{Frascati Phys. Ser.} \textbf{59}, 25 (2014) [arXiv:1409.2132 [hep-ex]]; \textit{Acta Phys. Polon. B} \textbf{46},
13 (2015) [arXiv:1501.04801 [hep-ex]].
\bibitem{BozenaActa} B. Jasi{\'n}ska \textit{et al., Acta Phys. Pol. B} \textbf{47}, (2016), these proceedings.
\bibitem{Bozena25}  R. Ferragut \textit{et. al, J. Phys.: Conf. Ser.} \textbf{225}, 012007 (2010).
\bibitem{Bozena26}   G. Consolati \textit{et al., Chem. Soc. Rev.} \textbf{42}, 3821 (2013).
\bibitem{Bozena28}   C. J. Edwardson \textit{et al., J. Phys.: Conf. Ser.} \textbf{262}, 012018 (2011).
\bibitem{Bozena29}   A. Kierys \textit{et al., Microporous Mesoporous Mater.} \textbf{179}, 104(2013).
\bibitem{Coleman}    P. Coleman, Positron Beams and Their Applications, World Scientific, (2000).
\bibitem{Bernreuther1981} W. Bernreuther, O. Nachtmann, \textit{Z. Phys. C} \textbf{11}, 235 (1981).
\bibitem{KleinNishina} O. Klein, T. Nishina, \textit{Z. Phys.} \textbf{52}, 853 (1929).
\bibitem{Evans} R. D. Evans, Corpuscles and Radiation in Matter II, Springer Berlin Haidelberg, 218-298 (1958).
\bibitem{Beatrix} B. Hiesmayr \textit{et al.,} in preparation.
\bibitem{Daria-ops} D. Kami{\'n}ska \textit{et al.,} in preparation.
\bibitem{Daria}     D. Kami{\'n}ska \textit{et al., Nukleonika} \textbf{60}, 729 (2015) \newline [arXiv:1509.01114 [physics.ins-det]].
\bibitem{m52} A. Rich, \textit{Rev. Mod. Phys.} \textbf{53}, 127 (1981).
\bibitem{m53} H. S. Mani, A. Rich,  \textit{Phys. Rev. D} \textbf{4}, 122 (1971).
\bibitem{m54} V. B. Berestetskii, E. M. Lifshitz, L. P. Pitaevskii, Relativistic Quantum Theory, Pergamon Press, 1971.
\bibitem{m55} O. Sausa \textit{et al., Acta Phys. Pol. A} \textbf{113}, 1517 (2008).
\bibitem{P3} P. Moskal, Patent number: WO2015028604, PCT/EP2014/068374.
\bibitem{mpatent1} P. Moskal, Patent number: WO2011008119-A2; PL388555-A1; US2012112079-A1; EP2454612-A2; JP2012533734-W.
\bibitem{EwelinaNukleonika} E. Kubicz \textit{et al., Nukleonika} \textbf{60}, 749 (2015) [arXiv:1509.00411 [q-bio.OT]].
\bibitem{AnnaActa} A. Wieczorek \textit{et al., Acta Phys. Polon. A} \textbf{127}, 1487 (2015) \newline [arXiv:1502.02901 [physics.ins-det]].
\bibitem{AnnaNukleonika} A. Wieczorek \textit{et al., Nukleonika} \textbf{60}, 777 (2015) \newline [arXiv:1508.06820 [physics.ins-det]].
  
\end{thebibliography}
\end{document}